\documentclass[twocolumn]{emulateapj}  
\usepackage[left]{lineno}
\usepackage{epstopdf}
\usepackage{ulem}
\usepackage{amssymb}
\usepackage{natbib}
\usepackage{times}
\usepackage{graphicx}
\usepackage[usenames,dvipsnames]{pstricks}

\usepackage{psfrag}
\usepackage{lipsum}
\usepackage{epstopdf}
\usepackage[colorlinks=true,linkcolor=blue,citecolor=blue]{hyperref}

\usepackage{adjustbox}

\newcommand{\msun}{\,$M_{\odot}$}	

\shorttitle{Follow-up observations of GW200115}
\shortauthors{Dichiara et al.}

\begin{document}

\title[DDOTI observations of GW200115]{Constraints on the electromagnetic counterpart of the Neutron Star Black Hole merger GW200115}

\author{
S.~Dichiara\altaffilmark{1,2,3}\email{sbd5667@psu.edu}, 
R.~L.~Becerra\altaffilmark{4},
E.~A.~Chase\altaffilmark{5,6},
E.~Troja\altaffilmark{1,2},
W.~H.~Lee\altaffilmark{7},
A.~M.~Watson\altaffilmark{7},
N.~R.~Butler\altaffilmark{8},
B.~O'Connor\altaffilmark{1,2,9,10},
M.~Pereyra\altaffilmark{11},
K.~O.~C.~L\'opez\altaffilmark{12},
A.~Y.~Lien\altaffilmark{13},
A.~Gottlieb~\altaffilmark{1,2},
 and
A.~S.~Kutyrev\altaffilmark{1,2}, 
\\
}
\affil{$^{1}$ Department of Astronomy, University of Maryland, College Park, MD 20742-4111, USA\\
$^{2}$ Astrophysics Science Division, NASA Goddard Space Flight Center, 8800 Greenbelt Road, Greenbelt, MD 20771, USA\\
$^{3}$ Department of Astronomy and Astrophysics, The Pennsylvania State University, 525 Davey Lab, University Park, PA 16802, USA \\
$^{4}$ Instituto de Ciencias Nucleares, Universidad Nacional Aut\'onoma de M\'exico, Apartado Postal 70-264, 04510 M\'exico, CDMX, Mexico\\
$^{5}$ Center for Theoretical Astrophysics, Los Alamos National Laboratory, Los Alamos, NM, 87545, USA\\
$^{6}$ Intelligence and Space Research Division, Los Alamos National Laboratory, Los Alamos, NM, 87545, USA\\
$^{7}$ Instituto de Astronom{\'\i}a, Universidad Nacional Aut\'onoma de M\'exico, Apartado Postal 70-264, 04510 M\'exico, CDMX, Mexico \\
$^{8}$ School of Earth and Space Exploration, Arizona State University, Tempe, AZ 85287, USA\\
$^{9}$ Department of Physics, The George Washington University, 725 21st Street NW, Washington, DC 20052, USA\\
$^{10}$ Astronomy, Physics and Statistics Institute of Sciences (APSIS), The George Washington University, Washington, DC 20052, USA\\
$^{11}$ CONACYT, Instituto de Astronom{\'\i}a, Universidad Nacional Aut\'onoma de M\'exico, 22860 Ensenada, BC, Mexico\\
$^{12}$ Facultad de Ciencias, Universidad Nacional Aut\'onoma de M\'exico, A. P. 70-543, 04510 D.F, Mexico \\
$^{13}$  University of Tampa, Department of Chemistry, Biochemistry, and Physics, 401 W. Kennedy Blvd, Tampa, FL 33606, USA\\
}

\begin{abstract}

We report the results of 
our follow-up campaign for the neutron star - black hole (NSBH) merger GW200115 detected during the O3 run of the Advanced LIGO and Advanced Virgo detectors.
We obtained wide-field observations with the Deca-Degree Optical Transient Imager (DDOTI)  covering $\sim$20\% of the total probability area down to a limiting magnitude of $w$=20.5 AB at $\sim$23~h after the merger. 
Our search for counterparts returns a single candidate (AT2020aeo), likely not associate to the merger. In total, only 25 sources of interest were identified by the community and later discarded as unrelated to the GW event. 
We compare our upper limits with the emission predicted by state-of-the-art kilonova simulations and disfavor high mass ejecta ($>$0.1\msun), indicating that the spin of the system is not particularly high. By combining our optical limits with gamma-ray constraints from {\it Swift} and {\it Fermi}, we  disfavor the presence of a standard short duration burst for viewing angles $\lesssim$15 deg from the jet axis. 
Our conclusions are however limited by the large localization region of this GW event,  
and accurate prompt positions remain crucial to improving the efficiency of follow-up efforts. 

\end{abstract}



\section{Introduction}
\label{sec:introduction}

Compact binary mergers composed of two neutron stars (NSs) or a NS and a stellar mass black hole (BH) have long been suspected of being the possible progenitors of at least a class of gamma-ray bursts (GRBs; 
\citealt{Eichler89,Kluzniak98,Janka99,Rosswog04,Faber06})
and of being relevant to the production of r-process elements \citep{Lattimer74}. 
Depending on the BH mass, its spin and the NS compactness, 
the close encounter of the two compact objects could lead to the tidal disruption of the NS \citep[e.g.][for a recent exploration of the parameter space]{Foucart18}. 
When this happens, a hot ($T$\,$\sim$\,10 MeV) and massive (up to $\sim$0.3\,\msun) accretion disc is formed around the BH, likely providing the energy source for a GRB \citep{Popham99,Narayan01,Lee05}. A large amount of neutron-rich matter is also ejected in the form of tidal tails and disc outflows, and its radioactive decay could power a bright kilonova emission \citep{Li98,Roberts11,Kyutoku15,Fernandez20}.
Therefore, the expectation is that NSBH mergers should be characterized by  electromagnetic (EM) counterparts similar to binary NS mergers such as GW170817
\citep[e.g.][]{Abbott17MMA}.  
However, the presence of a primary BH with a significantly higher mass and a definite event horizon, rather than a NS, unavoidably changes the merger dynamics and is likely to  leave distinguishable imprints in the EM radiation, such as the color and luminosity of the kilonova or a long-lasting hard X-ray emission \citep{Norris06,Troja08,Barbieri20,Dichiara21}. 
An outstanding issue is whether in a NSBH event the dynamics will allow for full disruption of the NS and the formation of an accretion disk that will both release gravitational binding energy into a relativistic outflow to power the GRB and eject neutron-rich matter leading to a kilonova, or if the NS will be accreted whole leading to little or no counterpart EM emission.  

Although NSBH binaries and their mergers were a long-standing prediction of stellar evolution models 
\citep{Yungelson1998,Pfahl2005},
proof of their existence remained elusive until recently. 
The advent of gravitational wave (GW) astronomy provided us with a new avenue to discover binaries of compact objects: over 40 binary black holes (BBHs) mergers, and two NS mergers were identified  through GW searches \citep{GWTC2,GWTC1}. In addition, several NSBH candidates were found during the third run of Advanced LIGO and Virgo observations.

\begin{figure*}[htp]
\centering
 \includegraphics[width=0.95\textwidth]{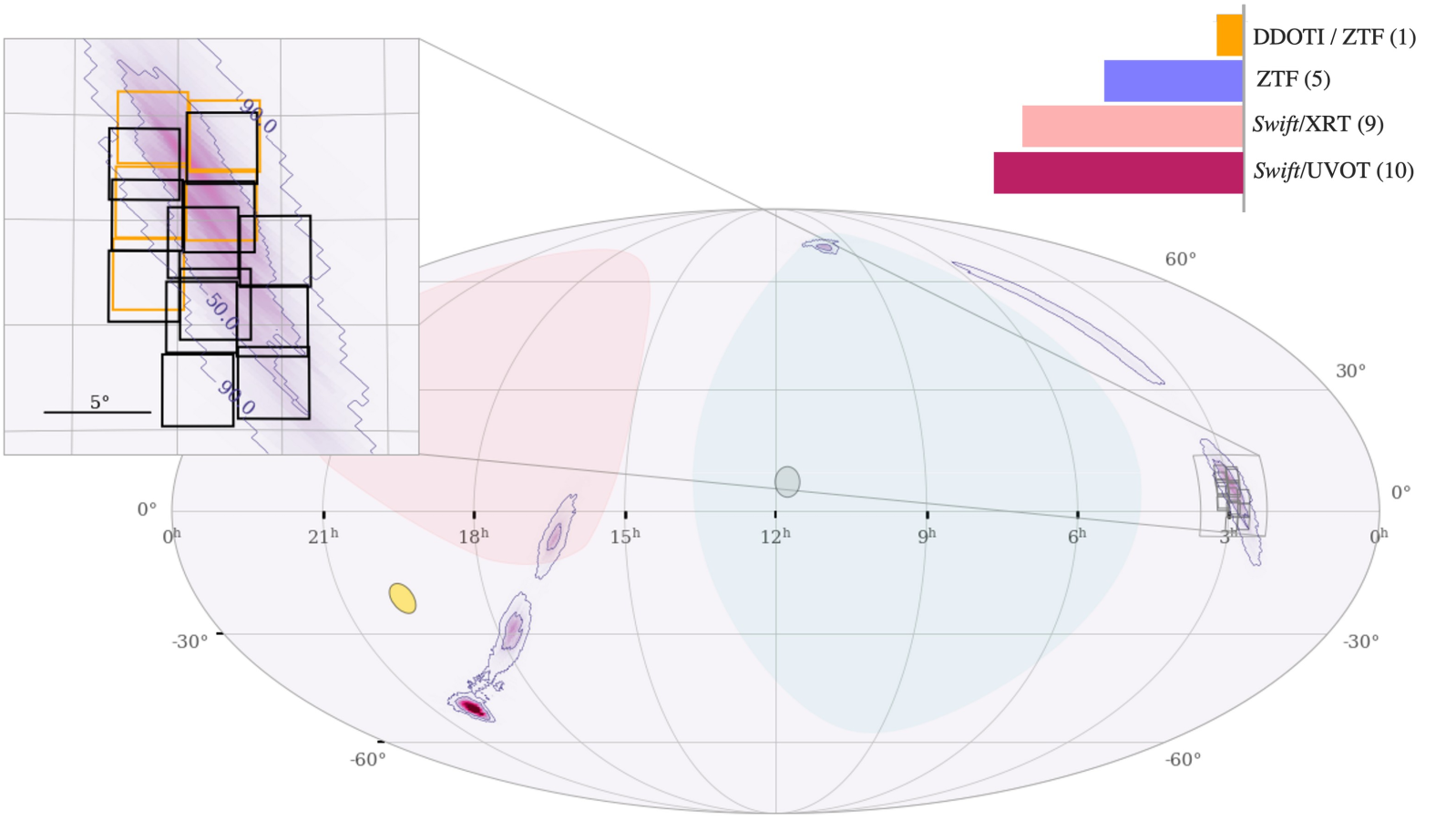}
 \caption{The LIGO/Virgo LALInference skymap of GW200115
 showing two probability lobes.
The BAT field of view at the time of the merger is shown by the orange shaded area. 
The Fermi/GBM field of view is all the sky but the blue region.
 DDOTI fields are shown as grey squares. The Sun and Moon (at the moment of DDOTI observations) are indicated by yellow and grey circles, respectively.
 In the zoom-in, blue contours illustrate the probability contours of 50\% and 90\%. 
 Orange and black squares illustrate the fields observed by DDOTI in the first and second night, respectively. 
 At the top, the bar chart reports the candidates identified by DDOTI, ZTF \citep{Anand2021}, and \textit{Swift} \citep{Page20,Oates21}.}
 \label{fig:observations}
\end{figure*}

The first two candidate NSBH mergers were GW190426
and GW190814 \citep{LVC190814}. The former has a low statistical significance and cannot be confidently associated with an astrophysical event, the latter has a secondary mass of $\approx$2.5\msun\ which does not allow for an unambiguous classification as either NS or BH. None of these GW signals was associated to an EM counterpart 
\citep{Andreoni20,Ackley20,Dobie20,Page20,Thakur20,deWet21,Watson20,Becerra21,Oates21}.
In January 2020, two more NSBH candidates were discovered with the advanced LIGO and Virgo interferometers: GW200105, detected with high significance only by LIGO Livingston and therefore poorly localized to a 90\% credible region of 7200 deg$^2$, and GW200115, seen by all three GW detectors and localized within a 90\% credible region of 600 deg$^2$ \citep{Abbott2021}. Their component masses are approximately 8.9\msun~and 1.9\msun~for GW200105, and 5.7\msun~and 1.5\msun~for GW200115, which confidently place them in the NSBH range. 
In all these cases, the BH spin is estimated to be low or anti-aligned to the orbital angular momentum, which reduces the chance of a NS tidal disruption, hence of  a bright EM counterpart. 
Furthermore, due to their high mass ratio, GW190814 and GW200105 are often considered as either plunging events, in which the NS was entirely swallowed by the BH, or low-mass BBHs. 

In this paper, we focus on GW200115 which, due its relatively contained localization and low mass ratio, is thus far the best candidate to constrain the EM counterparts of a NSBH merger.
Different satellites, including {\it Swift}, {\it Fermi}, {\it AGILE}, {\it MAXI} and {\it CALET}, covered part of the probability region of this GW event but no counterparts were detected at high energies.
Six candidates were discovered during the optical follow-up campaign of this GW event \citep{Anand2021}.
It is interesting to note that the number of optical candidates reported 
for GW200115 is much lower than the ones found in the case of GW190814, 
for which 85 candidates were reported within the 19~deg$^2$
localization region \citep[see][]{Thakur20}. 
Most of the candidate counterparts were identified by \textit{Swift}:
nine XRT sources, classified as possible ``sources of interest'' \citep[Rank 2; ][]{Page20}, 
and ten faint UVOT sources \citep{Oates21}. The majority of these were later associated to AGN activity. 

 In \S\ref{sec:observations} we present the data analysis results of our observing campaign of GW200115
with the Deca Degree Optical Transient Imager \citep[DDOTI;][]{Watson16}, the wide-field optical imager located at the Observatorio Astronómico Nacional (OAN) on the Sierra de San Pedro Mártir in Mexico.
In \S\ref{sec:constraints} we outline different constraints from our observations. Finally, in \S\ref{sec:summary} summarize our results and discuss their implications. \\

\section{Observations and Data Analysis}
\label{sec:observations}

On January 15, 2020 at 04:23:09.7 UT the GW network
composed of LIGO Hanford (H1), LIGO
Livingston (L1), and the Virgo Observatory (V1) 
triggered on the gravitational wave transient S200115j \citep{2020GCN.26759....1L}.
Preliminary analysis classified it as a MassGap event, 
that is a merger in which at least one of the two objects has a mass in the range $\approx$3-5 M$_{\sun}$. The lighter object was identified as a likely NS. 

The GW transient was initially localized within a sky area of 
$\approx$900~deg$^2$ (90\% credible region) at a distance of
$\approx$330 Mpc. The probability map presents two main lobes,  one peaking in the Southern hemisphere at RA, Dec (J2000) = 298.50, -49.60 deg, 
and the other one peaking in the Northern hemisphere at RA, Dec (J2000) = 42.75, 5.70 deg. 

DDOTI received the alert during the observing night. 
Its automatic response generated a pointing centred on the peak of the 2D probability map visible from the Northern hemisphere (see Figure~\ref{fig:observations}). 
Observations started 1.3 h after the GW trigger and monitored the field for a total exposure of 32 min, carried out in poor weather conditions (clouds).  
These images reach a 10\,$\sigma$ point-source sensitivity of about 17.3-18.4 AB mag. 
Since one of the detectors was out-of-focus, only five of them were used to cover the GW region
with a total field of view of 59\,deg$^2$. This corresponds to 11\% of the LALInference probability map. 
Due to the large localization, the southern location of the probability peak, and the bad weather, no tiling of the GW region was planned. 

On the second night, observations were carried out with all six detectors and in better weather conditions. 
A larger sky area of 115 deg$^2$ (19\% of the LALInference Map) was covered down to a 10~$\sigma$ limiting magnitude of 20.1 AB mag. 

We used observations taken on August 20 and August 19, 2021 (e.g. $\sim$1.6 years after  the burst) as a template for the image subtraction. The only candidate identified in the residual images after excluding fast-moving solar system objects and image artifacts is ZTF20aagjqxg/AT2020aeo \citep{Anand2021}. Using PSF photometry we measure a magnitude of $w$=20.12$\pm$0.28 AB mag for this source. This transient was not promptly detected from the automatic pipeline because of two different factors: 1) the source was at the edge of the camera where the instrument's sensitivity is lower; 2) the source is weak. It is below the 10~$\sigma$ significance required to trigger the detection and it can only be recovered using image subtraction. This source was ruled out as a possible counterpart due to its slowly rising light curve \citep{Anand2021}.
Although we found that the temporal evolution could be consistent with the one expected from an off-axis jet seen at small viewing angles ($\theta_{v}$ $<$ 15 deg)
with a large energetic ($E_{\gamma,iso}$\,$>$\,10$^{52}$~erg), the event would be likely associated with a prompt emission gamma-rays signal. Therefore, the combination of our optical limits optical with the lack of detected gamma-rays emission disfavor the off-axis model.

No other reliable counterpart was found in the residual images down to a 7~$\sigma$ limit of $\approx$17.7 AB mag and $\approx$20.5 AB mag at 1.3 hours and 22 hours from the gravitational wave trigger, respectively. These values roughly correspond to 3~$\sigma$ limits post-trial, taking into account the number of independent elements in the field.
In addition to AT2020aeo, detected by ZTF and DDOTI, five other optical candidates were reported by \cite{Anand2021}. All of them lied outside the field covered by DDOTI, and were discarded based on either spectroscopic or photometric observations.

The number of candidate counterparts is small compared with the ones discovered during the follow-up campaigns of other events (e.g. GW 190425 or GW 190814).
Several factors might have contributed to decrease the efficiency of the search: the shallower depth of the observing campaign, which reached a limit of $\approx$21 AB mag for GW200115 and $\approx$23 AB mag for GW190814, the larger sky area searched for transients, as well as poorer weather conditions. 
In Figure~\ref{fig:observations} we summarize the results of the follow-up campaign, showing the observed fields and the candidates discovered by different instruments

\section{Constraints to the electromagnetic counterparts}
\label{sec:constraints}

\subsection{GRB Prompt Emission}

At the time of the merger, $\sim$12\% of the GW probability region was within the field of view of the \textit{Swift} Burst Alert Telescope (BAT; Figure~\ref{fig:observations}) with partial coding $<$60\%. 
We use the BAT data to constrain any possible prompt gamma-ray emission from GW200115. 
The prompt phase of short GRBs is characterized by three main features:
a) a precursor, visible in $<$15\% of the events 
\citep[e.g. ][]{Troja10},
b) a main peak of short duration and hard spectrum and 
c) a temporally extended emission with soft spectrum \citep{Norris06}.  
The latter component is visible in $<$20\% of the events, although might be present in a larger fraction of events and be undetected due to instrumental effects \citep{Dichiara21}.

Precursors might precede the merger by a few seconds. 
Their origin is not well understood but is commonly interpreted within the framework of binary NS interactions \citep[e.g.][]{Tsang2012}, and may not be present in NSBH mergers.

The main short-duration GRB is produced by a highly-relativistic jet launched by the merger remnant and is therefore expected to occur right after the merger event.  Due to the beamed geometry of the outflow, it is visible  only to observers close to the jet's axis. 
No transient was detected by BAT within 4~s of the GW trigger down to a 3 $\sigma$ upper limit of $\approx$ 9$\times$10$^{-8}$ erg cm$^{2}$ s$^{-1}$,
corresponding to an isotropic equivalent luminosity of $\approx$10$^{48}$ erg s$^{-1}$ at a distance of 300 Mpc. 
This is below the typical luminosity of cosmological short GRBs 
($\sim 10^{51}$ erg s$^{-1}$; \citealt{Lien2016})  
and strongly disfavors the presence of an on-axis explosion, whereas off-axis jets cannot be ruled out. 
The GW localization was only partially covered by the BAT, but fell within the field of view of the  {\it Fermi} Gamma-Ray Burst Monitor (GBM) that covered $\sim$96\% of the probability region including also the field observed by DDOTI. Also in this case, the GBM did not detect any short duration signal around the time of the GW trigger down to a 3 $\sigma$ flux limit of $\approx$ 3$\times$10$^{-7}$ erg cm$^{2}$ s$^{-1}$ \citep{2020GCN.26774....1G}. Assuming the luminosity distance derived from the study of the gravitational wave signal (300 Mpc) this corresponds to an isotropic equivalent luminosity limit of $4 \times 10^{48}$ erg s$^{-1}$, which rules out the presence of a typical short GRB (the only short burst with a luminosity lower than this value is GRB 170817A).

The short GRB peak is sometimes followed by a prolonged tail of emission. A typical example is GRB~050724A \citep{Barthelmy05}, a short burst followed by a tail of spectrally soft emission with a luminosity of $4 \times 10^{48}$ erg s$^{-1}$ and duration of $\approx$100 s.
In some models, this extended emission is linked to NSBH mergers and the fallback accretion of NS matter on the central BH  \citep[e.g. ][]{Rosswog2007}. 
The outflow powering this emission is not necessarily characterized by the same degree of collimation of the initial GRB jet and, if wider, could be seen by an off-axis observer. 
Therefore, we searched for long-duration transients in the BAT rate light curves during the time interval 5-100 s after the GW trigger. 
Assuming a typical power-law spectrum with photon index $\Gamma$=1.8 \citep{Lien2016} we derive a 3 $\sigma$ upper limit of $\approx$7$\times$10$^{-6}$ erg~cm$^{2}$ on the fluence of the extended emission.
At such sensitivity, a soft tail similar to or brighter than GRB~050724A would be detectable up to $z\lesssim$0.35. 

\begin{figure}[t!]
\centering
 \includegraphics[scale=0.55]{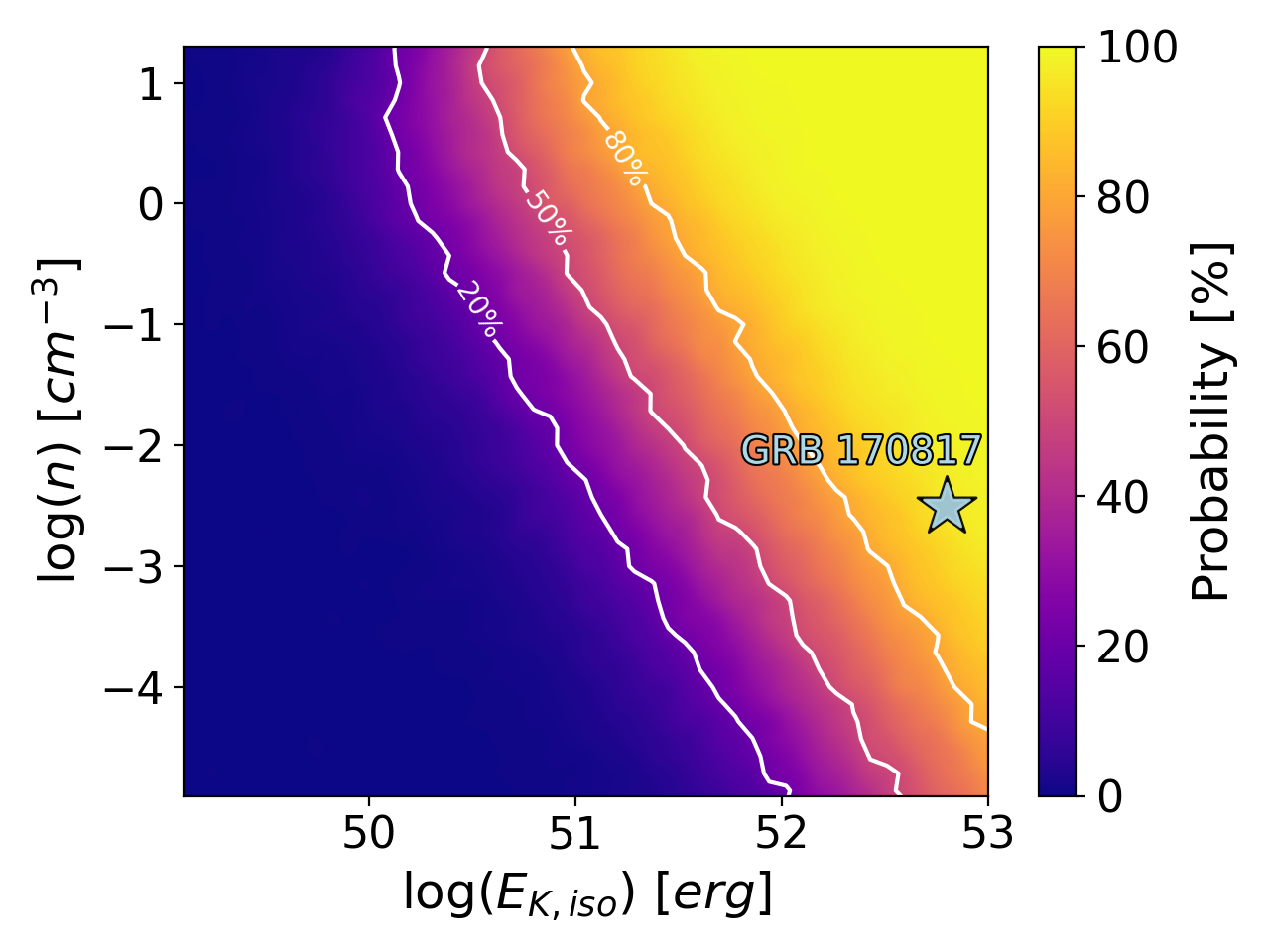}
  \includegraphics[scale=0.55]{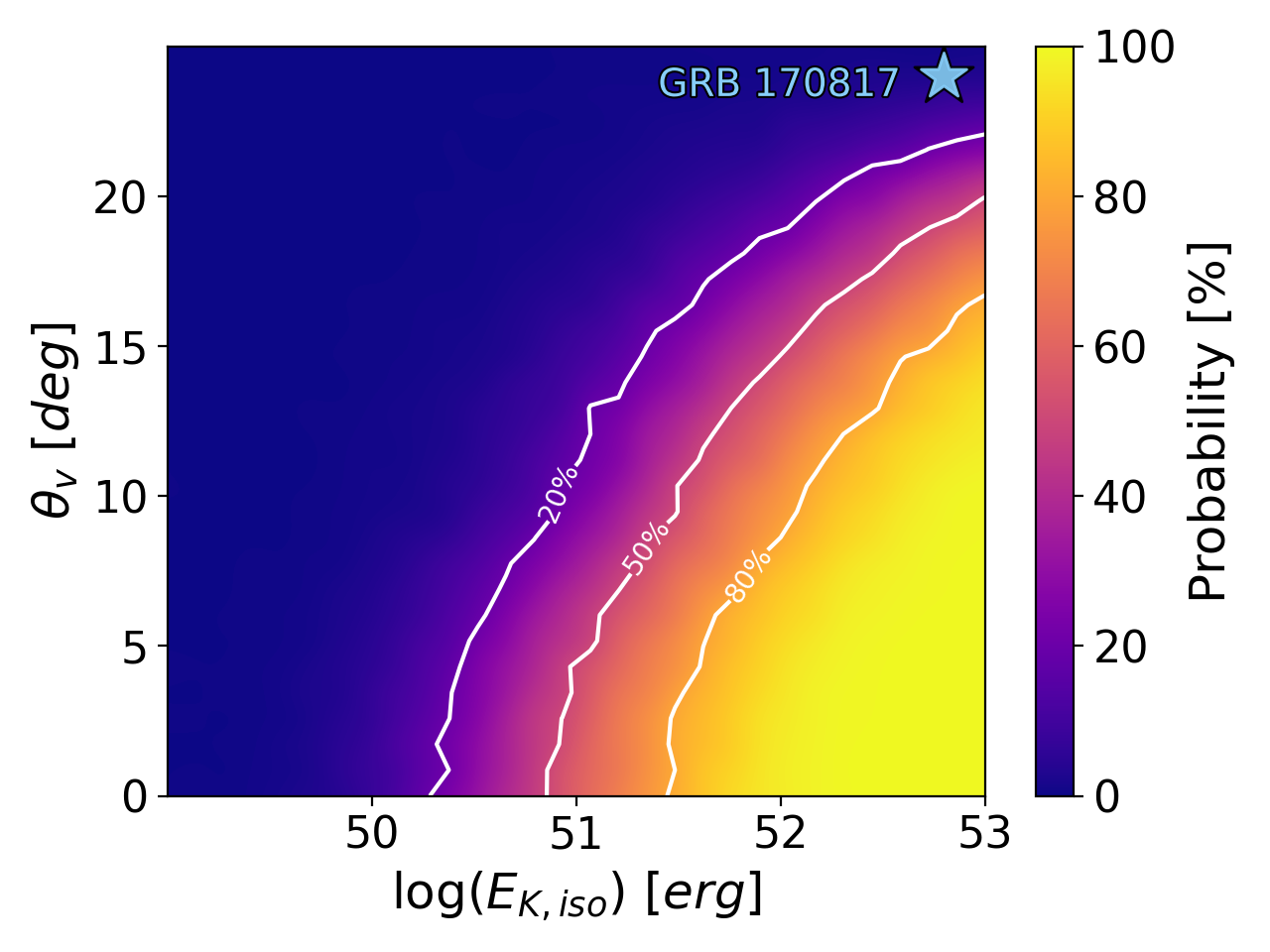}
 \caption{Top: Probability of detection for an on-axis short GRB assuming different densities for the external medium and total kinetic energy. The blue star shows the parameters derived for GRB~170817A \citep{Troja2019}.
 Bottom: Probability derived for different viewing angles and kinetic energy assuming a density of n=0.1 cm$^{-3}$.}
 \label{fig:sgrb_detProb}
\end{figure}

\subsection{GRB Afterglow}

If a successful GRB jet is launched by the merger remnant, its interaction with the surrounding medium will produce a broadband synchrotron radiation known as afterglow. 
The properties of the afterglow are determined by the energy of the explosion $E_K$, the density of the environment $n$, and the shock microphysics, described through the parameters $\epsilon_{B}$,
$\epsilon_{e}$, and $p$, which represent the fraction of energy that goes into the magnetic field, the fraction of energy that goes into the electrons, and the spectral index of the electron energy distribution $N(E) \propto E^{-p}$, respectively. 
An additional key parameter is the observer's viewing angle $\theta_v$, which can be aligned to the jet-axis (on-axis) or not (off-axis). 
In the former case, the observed afterglow emission reaches its peak luminosity soon after the GRB and rapidly fades away. 
In the latter case, the afterglow peak is delayed by several days or even months, and its peak luminosity quickly drops when moving away from the axis \citep[e.g. ][]{Ryan2020}.

We use DDOTI upper limits to constrain a possible afterglow emission following GW200115. 
A comparison to the observed light curves of short GRBs, shifted to a common distance of 300 Mpc (cf. Fig. 8 of \citealt{Thakur20}),  shows that our limits could rule out 50\% of the observed  events, as also found for GW190814 \citep{Thakur20,Watson20}. 

We also simulated a large set of optical afterglow light curves, 
representative of the sGRB population, using \textsc{afterglowpy} \citep{Ryan2020}. 
We used the distance posterior distributions of \cite{Abbott2021}, 
a log-normal distribution for $\epsilon_{e}$ centered at -1 with width $\sigma=0.3$ \citep{Beniamini2017}, and a log-uniform distribution for $\epsilon_{B}$ ranging between $-4$ and $-1$ \citep{Santana2014}. The spectral index $p$ was set to 2.3. 
We then created a grid of 100$\times$100 elements to take into account different values for the total kinetic energy and the density of the external medium ranging from $10^{49}$ to $10^{53}$ erg and from $10$ to $10^{-5}$\,cm$^{-3}$, respectively. 

The outcome of these simulations is summarized in Figure~\ref{fig:sgrb_detProb}.
For an on-axis observer ($\theta_v$=0), bright explosions ($E_{K,iso} > 10^{50}$ erg) in an ISM environment  with $n >$ 0.01 cm$^{-3}$ are likely to be detected (for short bursts the typical kinetic energy ranges between 10$^{51}$ erg and 10$^{54}$ erg).
Mergers in a rarefied environment ($n <$ 10$^{-4}$ cm$^{-3}$), such as those kicked out of their host galaxy, have instead a low probability to be detected via their afterglow, even if observed on-axis. 

We carried out a similar set of simulations for off-axis GRB explosions. 
In this case, we fixed the density to $n$=0.1 cm$^{-3}$ and varied the viewing angle from 0 to 25 deg. The probability of detection is shown in Figure~\ref{fig:sgrb_detProb}. 
According to the posterior distributions reported by \cite{Abbott2021}, 
the inclination angle of GW200115 has a probability of $\sim$ 10\% of being $\lesssim$15 deg. For this range of angles, our simulations show that the off-axis afterglow would have a high probability ($>$ 50\,\%) of being detected when the GRB energy is higher than $10^{51}$ erg.
Therefore, the combination of GW and EM constraints disfavors small ($\lesssim$15 deg) viewing angles. 

The sensitivity of our observations decreases for larger viewing angles. We can not detect events far from the jet axis (e.g. $\theta_{v}$ $>$ 20 deg) even assuming high energy explosions ($\sim 10^{53}$ erg). For example, an explosion similar to GRB~170817A would not have been detected.

\subsection{Kilonova emission}

Compact object mergers can drive outflows of neutron-rich material expanding at sub-relativistic speeds. 
The radioactive decay of this ejecta powers a luminous optical and near-infrared transient, known as kilonova or macronova. 
We offer constraints on a potential kilonova associated with GW200115 inside the field covered by the DDOTI observations, by comparing our deepest upper limit of $w$\,$>$\,20.5~AB~mag to the Los Alamos National Laboratory grid of kilonova simulations \citep{Wollaeger2021}.
\cite{Wollaeger2021} renders each of the 900 simulated kilonovae in 54 disparate viewing angles, distributed uniformly in cosine and each subtending an equal solid angle. In total, this results in 48,600 kilonova models to compare to observations.
This grid includes 900 multi-dimensional radiative transfer simulations, which jointly evolve dynamical (lanthanide-rich) and wind (lanthanide-poor) ejecta components, spanning a diverse set of ejecta masses, velocities, morphologies, and compositions (see \citealt{Wollaeger2021} and references therein).
These state-of-the-art simulations rely on updated lanthanide opacities from \citet{Fontes2020}. 
Although we can exclude kilonovae spanning a broad range of viewing angles, the majority of simulated kilonovae excluded by our observations correspond to events observed near the polar axis. Similar conclusions were obtained by \cite{Anand2021}, although they only rule out near-polar viewing angles and make no constraints on edge-on inclinations (see their Extended Data Figure 4). These differences are due to the different ejecta morphology used in the simulations.

\begin{figure}[t!]
\centering
 \includegraphics[width=0.48\textwidth]{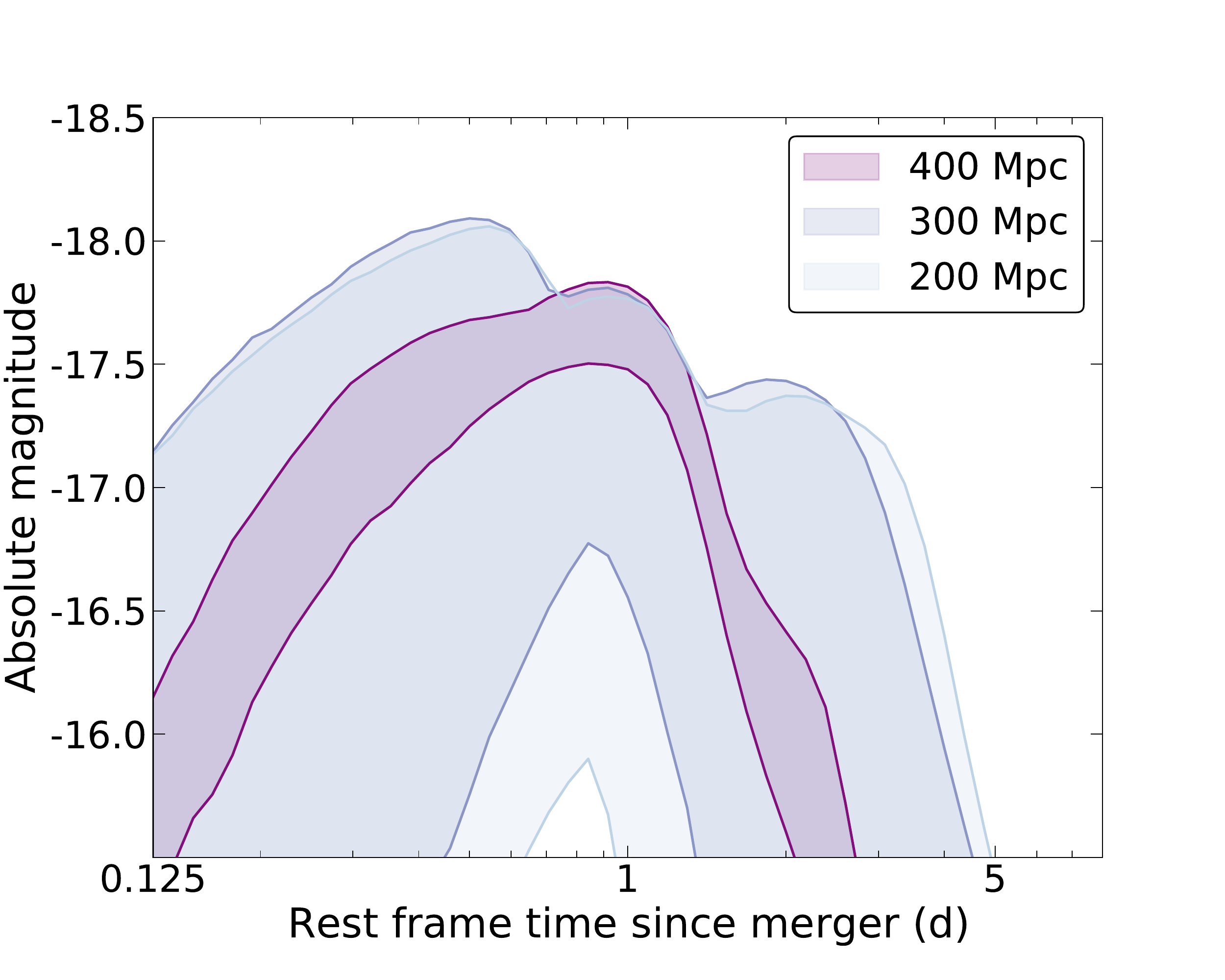}
 \caption{
 Range of simulated kilonova lightcurves \citep{Wollaeger2021} excluded by upper limits. All lightcurves are presented in the rest frame, for three distances within GW200115's 90\% credible interval on luminosity distance. Colored regions indicate the maximum range on absolute magnitude excluded by upper limits. Smaller luminosity distances correspond to deeper limiting magnitudes and, thus, an increased fraction of ruled out simulated lightcurves.}
 \label{fig:kilonova}
\end{figure}

Simulated kilonova spectra are converted to lightcurves in DDOTI's observer frame for six luminosity distances within the 90\% credible interval on GW200115's inferred distance: 200, 250, 300, 350, 400, and 450~Mpc. 
At each distance, we determine the fraction of simulated light curves excluded by an upper limit of 20.5 AB mag between 22--23.3 hours post-merger (observer-frame).
Figure~\ref{fig:kilonova} presents the range of rest-frame lightcurves ruled out by the upper limit for three distances.
The upper limit provides negligible constraints for 450~Mpc, the largest distance we consider. 
Larger fractions of simulated kilonovae curves are ruled out as distance decreases, with 0.94\%, 3.8\%, 5.3\%, 9.5\%, and 18\% excluded at luminosity distances of 400, 350, 300, 250, and 200~Mpc, respectively.
Excluded kilonovae broadly correspond to high wind ejecta masses ($>$ 0.03\,\msun); at the lowest distance considered (200~Mpc), over 65\% of simulated kilonovae with wind ejecta masses of 0.1\,\msun~are ruled out by our observations.
For comparison, the vast majority of ruled out lightcurves at higher distances also correspond to high wind ejecta mass, with 40\%, 27\%, 19\%, and 5\% of simulated kilonovae with wind ejecta masses of 0.1\,\msun~ruled out for luminosity distances of 250, 300, 350, and 400~Mpc, respectively.
We note that stronger constraints could have been placed on a potential kilonova associated with this source if additional observations were made, preferably between 6 and 18 hours post-merger \citep{Chase2021}.

\begin{figure}
\centering
 \includegraphics[width=0.45\textwidth]{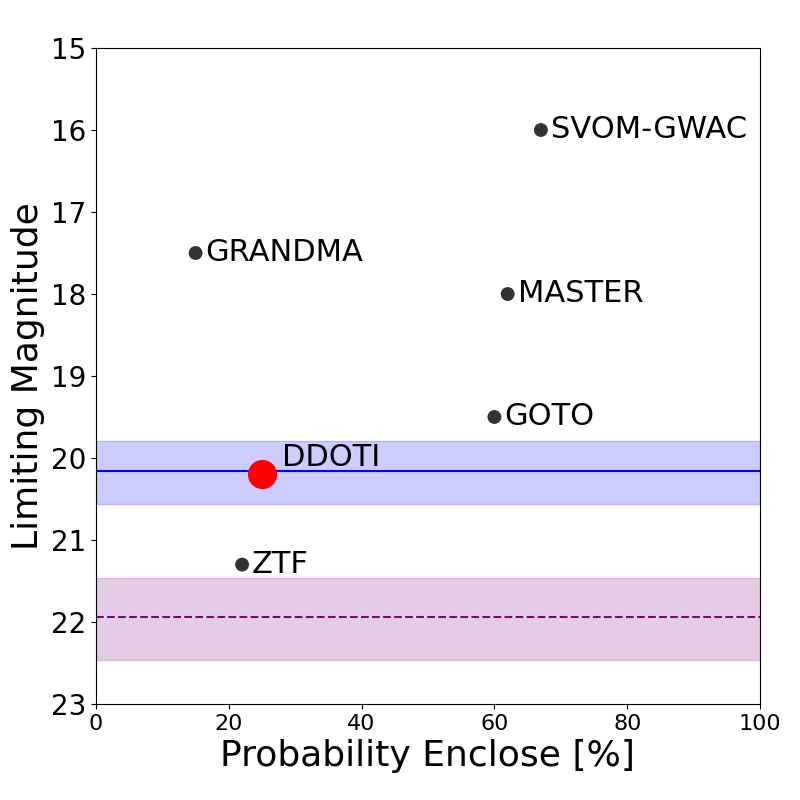}
 \caption{Comparison of limiting magnitudes and probabilities covered for GW200115 by different facilities. The purple dashed line and shaded region represents an event similar to AT2017gfo at a luminosity distance of 300 Mpc at $T+22$~h. The blue line shows the magnitudes expected for kilonovae models of \citet{Barbieri20} for NSBH merger with a maximally spinning BH at $T+22$~h. The shaded areas reflect the 1-$\sigma$ uncertainty in the source distance scale.}
 \label{fig:comparison}
\end{figure}

Our results are summarized in Figure~\ref{fig:comparison}, which reports the upper limits from DDOTI and other wide-field optical facilities. 
They are compared with the expected magnitude of a kilonova from 
a NSBH merger with a maximally spinning BH \citep{Barbieri20} 
and the magnitude of AT2017gfo, rescaled at a distance of 300 Mpc and at $22$~h after the merger.
For GW200115, the analysis of the gravitational wave signal does not allow us to put strong constraints on the BH spin magnitude \citep{Abbott2021}, however our EM upper limits disfavor high values of the spin parameter that would lead to high wind ejecta masses. 
We caution that this result is only valid within the field covered by DDOTI observations, however it shows the potential of joint EM-GW analysis for well-localized sources.

\section{Summary}
\label{sec:summary}

We used the EM observations of the gravitational wave event GW200115 to place constraints on the physical parameters of the NSBH merger and its ejecta. 

Combined optical and gamma-ray observations allow us to disfavor the presence of a short GRB viewed on-axis. 
Fist, the optical limits disfavor a GRB event with isotropic kinetic energy $E_{K,iso}$\,$\gtrsim$\,10$^{50}$~erg and a circumburst density $n$\,$\gtrsim$ 0.01 cm$^{-3}$, typical of an ISM environment. They also disfavor an event with a small viewing angle, $\lesssim$15 deg, although only for energies $E_{K,iso}$\,$\gtrsim$\,10$^{51}$~erg and densities $n$\,$\gtrsim$0.1~cm$^{-3}$. 
By adding the gamma-ray information, constraints are even tighter and exclude an event with 
$L_{\gamma,iso}$\,$\gtrsim$\,10$^{48}$~erg~s$^{-1}$. In the BAT data, which only covers 12\% of the GW probability map, we do not find any evidence for temporally extended emission following the merger, and  can rule out a component similar to the one detected in GRB~050724A up to $z\lesssim$0.35.

Due to the distance to this event, only a fraction of the kilonova models could be constrained.  A sizable portion (up to 65\%) of the simulated kilonova light curves can be excluded when considering a high wind ejecta mass $M>$0.1\,$M_{\odot}$, 
which is to be expected in cases of a maximally spinning BH. 
However, due to the limited sky coverage of DDOTI's observations ($\approx$20\%), 
no strong conclusion can be placed. 

Given the number of candidate NSBH merger events detected through gravitational waves to date and the absence of EM counterpart emission, it is clear that exploiting them to better understand their viability as sGRB progenitors and r-process nucleosynthesis production sites will require significantly more efficient follow-up both in terms of position determination and prompt deep multi-band imaging. This will be particularly relevant with the increased sensitivity expected during the coming O4 observing run.  

\section*{ACKNOWLEDGEMENTS}
We thank the staff of the Observatorio Astron\'omico Nacional.
DDOTI is funded by CONACyT (LN 260369, LN 271117, and 277901), the University of Maryland (NNX17AK54G), and the Universidad Nacional Autónoma de México (CIC and DGAPA/PAPIIT IG100414, IT102715, AG100317, and IN109418, AG100820, IN105921) and is operated and maintained by the Observatorio Astron\'omico Nacional and the Instituto de Astronom\'ia of the Universidad Nacional Aut\'onoma de México.  RLB acknowledges support from the DGAPA/UNAM IG100820.
Part of this work was performed at the Aspen Center for Physics. The Aspen Center for Physics is supported by National Science Foundation grant PHY-1607611.
EAC's contributions were supported by the US Department of Energy through the Los Alamos National Laboratory.
Los Alamos National Laboratory is operated by Triad National Security, LLC, for the
National Nuclear Security Administration of U.S.\ Department of Energy (Contract No.\ 89233218CNA000001).
This research used resources provided by the Los Alamos National Laboratory Institutional Computing
Program, which is supported by the U.S. Department of Energy National Nuclear Security Administration
under Contract No.\ 89233218CNA000001.
The work was also partially supported  by  the  National  Aeronautics  and  Space  Administration through grant 80NSSC18K0429 issued through the Astrophysics Data Analysis Program and the National Science Foundation grant 2108950.





\label{lastpage}

\end{document}